\documentclass[double]{article}
\usepackage{times}
\usepackage{algorithm}
\usepackage{algorithmic}
\usepackage{wrapfig}
\usepackage{verbatim}
\usepackage{amsmath}
\usepackage{graphicx}
\usepackage{subcaption}
\usepackage[active]{srcltx}
\usepackage{color}
\usepackage{lineno}
\usepackage{dirtytalk}
\usepackage{amsthm}
\usepackage{authblk}
\usepackage{listings}

\usepackage{epsfig,graphicx,amsfonts}
\usepackage{amsmath,amsthm,amssymb}
\usepackage{xcolor}

\usepackage{adjustbox}
\usepackage{multirow}
\usepackage{setspace}
\usepackage{hyperref}
\usepackage{comment}

\long\def\comment #1\commentend{}

\begin{document}

\title{\Large Investigation Toward The Economic Feasibility of Personalized Medicine For Healthcare Service Providers: The Case of Bladder Cancer}
\author{Elizaveta Savchenko$^{1}$, Svetlana Bunimovich-Mendrazitsky$^{1*}$\\
\(^1\) Department of Mathematics, Ariel University, Ariel, Israel\\
\(*\) Corresponding author: svetlanabu@ariel.ac.il

}

\date{}

\maketitle 

\begin{abstract}
In today's complex healthcare landscape, the pursuit of delivering optimal patient care while navigating intricate economic dynamics poses a significant challenge for healthcare service providers (HSPs). In this already complex dynamics, the emergence of clinically promising personalized medicine based treatment aims to revolutionize medicine. While personalized medicine holds tremendous potential for enhancing therapeutic outcomes, its integration within resource-constrained HSPs presents formidable challenges. In this study, we investigate the economic feasibility of implementing personalized medicine. The central objective is to strike a balance between catering to individual patient needs and making economically viable decisions. Unlike conventional binary approaches to personalized treatment, we propose a more nuanced perspective by treating personalization as a spectrum. This approach allows for greater flexibility in decision-making and resource allocation. To this end, we propose a mathematical framework to investigate our proposal, focusing on Bladder Cancer (BC) as a case study. Our results show that while it is feasible to introduce personalized medicine, a highly efficient but highly expensive one would be short-lived relative to its less effective but cheaper alternative as the latter can be provided to a larger cohort of patients, optimizing the HSP's objective better. \\ \\

\noindent
\textbf{Keywords:} Personalized medicine; healthcare economics; patient-centric care; resource-constrained healthcare; healthcare systems management.

\end{abstract}

\maketitle \thispagestyle{empty}

\pagestyle{myheadings} \markboth{Draft:  \today}{Draft:  \today}
\setcounter{page}{1}

\section{Introduction}
\label{sec:introduction}
Hospitals and other healthcare service providers (HSPs) constantly endeavor to provide the highest quality of care to patients in a complex business and economic context \cite{start_1,start_2}. Personalized medicine emerging as a promising approach to achieving this goal since it tailors treatments to individual patients based on their unique properties such as genetic makeup \cite{reviewer_ask}, lifestyle \cite{teddy_bcg_suger}, socio-demographic status \cite{liza_phd_1}, offering the potential to optimize therapeutic outcomes and minimize adverse effects \cite{teddy_cells}. However, the implementation of personalized medicine in resource-constrained hospitals poses significant challenges \cite{journal_2,intro_1,intro_2}. As patients increasingly seek the most personalized and effective treatments, hospitals must grapple with limited resources and the need to make economically viable decisions. This setup implies a counter-intuitive scenario where the objective of the patient does not align with the HSP while both wish to make the patient healthy again as quickly as possible \cite{paitent_happy}.

The concept of personalized medicine has garnered extensive attention in both academia and the healthcare industry \cite{per_1,per_2,per_3}. Previous studies have demonstrated the potential of personalized medicine in enhancing treatment efficacy and patient satisfaction \cite{per_happy_1,per_happy_2}. Researchers have explored various approaches, including pharmacogenomics, biomarker analysis, and decision-support systems, to identify optimal treatment strategies for individual patients \cite{per_method_1,per_method_2,per_method_3}. That said, most of these works focus on the clinical and patient levels, ignoring the operational and economic burden associated with different levels of personalized medicine. For example, if some illness does not require some text for the commonly used treatment protocol while the personalized alternative does require this test, the usage of the personalized treatment protocol causes additional expenses in the form of more tests. 

Indeed, there is a growing body of literature about the efficient usage of resources in the healthcare sphere, ranging from a single patient to an entire hospital \cite{teddy_hospitals,ml_example_1,ml_example_2,ml_example_3}. However, economic evaluations of personalized medicine's economic impact have been mostly neglected. Hence, our understanding of the cost-effectiveness implementation of personalized treatment which ensures their practicality in an economically-driven real-world scenario is still lacking. 

One way to tackle this challenge is using an economical-mathematical model together with data science methods, as these have shown to be powerful tools in similar tasks \cite{eco_good_1,eco_good_2,eco_good_3,eco_good_4}. In practice, data science has emerged as a powerful tool in healthcare, revolutionizing the way medical decisions are made and HSPs are managed \cite{teddy_hospitals,intro_7,intro_8}. Data-driven approaches allow hospitals to identify subpopulations that would benefit the most from personalized medicine, thus addressing the challenge of resource scarcity \cite{model_intro_4,cancer_intro_3,intro_6}. 

In this work, we propose a comprehensive framework that addresses the economic feasibility of personalized medicine in resource-constrained healthcare settings. Our approach seeks to strike a balance between catering to individual patient needs and making cost-effective decisions on a broader scale. To this end, we proposed a novel mathematical model and its implementation as a computer simulation that aims to identify patient cohorts that are most likely to benefit from personalized treatments, as well as allow HSPs to choose between several levels of personalization in order to optimize both a clinical and economical objective. In order to investigate the model, we focused on bladder cancer (BC) disease, simulating (pseudo-)realistic scenarios. The novelty of this work lies in treating personalized treatment as a scale rather than a binary option. 

The remainder of this paper is structured as follows: In Section \ref{sec:related_work}, we provide an in-depth review of the existing literature on personalized medicine, highlighting relevant economic evaluations and challenges in resource-constrained settings. In addition, a review of treatment configurations for BC is presented. Section \ref{sec:model} outlines the formal model definition and its implementation as a computer simulation. Section \ref{sec:results} presents the results of our study. Finally, Section \ref{sec:discussion} concludes our findings and discusses the potential impact of personalized medicine on the healthcare domain while also providing suggestions for future work.

\section{Related Work}
\label{sec:related_work}
Like clinical treatment itself, personalized treatment can be seen as a spectrum as different methods require different amounts and versatility of data and promise different levels of outcome improvement as a result. In parallel, the extra effort associated with such data gathering is an economic concern that also divides into several sub-categories. In this section, we present the recent works on these two fronts with a focus on their interaction with our work. 

\subsection{Personalized medicine}
The application of personalized medicine in healthcare has garnered significant interest due to its potential to improve patient outcomes by tailoring treatments to individual characteristics \cite{rw_2_1,rw_2_2}. Numerous studies have explored the usage of personalized medicine in various medical conditions, providing evidence of its effectiveness and benefits \cite{rw_2_3,rw_2_4,rw_2_5}. Ce et al. \cite{rw_2_6} provided a detailed overview of the usage of personalized medicine for brain tumor imaging. The authors show that personalization can be done in the diagnosis, treatment protocol decision, and even post-treatment check-ups. For instance, Mzoughi et al. \cite{rw_2_7} proposed a conventional neural network-based model which utilized the whole volumetric T1 contrast-enhancement MRI sequence for MRI gliomas brain tumor classification. The authors analyze the performance of their model, showing it has great potential to improve the decision-making process of clinicians. Van Nettern et al. \cite{rw_2_8} investigated the future of personalized treatment for diabetic foot ulcer prevention, showing that using clinical and treatment clinical outcomes of similar patients is able to improve the treatment protocol patients obtain as both diagnosis conditions and treatment protocols are influenced by these factors. Yaniv-Rosenfeld et al. \cite{liza_before_phd} proposed a deep-learning-based model for BCG and IL-2 injections for BC immunotherapy treatment. The authors show that the personalized treatment protocol, which requires more often tests, outperforms the generic treatment protocol with a 12 percent higher success rate. In a different work, Yaniv-Rosenfeld et al. \cite{amit_new} show that using socio-demographic data, mental health professionals can better estimate the stay duration required by borderline personality disorder, which should improve their overall treatment, according to the authors. In this setting, the personalization requires one to fulfill a short personal questionnaire which is usually part of the administration process anyway, causing small to no additional effort to the HSPs. 

\subsection{Clinical resource allocation in healthcare settings}
Optimizing resource allocation in healthcare is crucial to ensure the effective and efficient delivery of healthcare services, especially in settings with limited resources \cite{rw_3_1,rw_3_2,rw_3_3}. Recently, multiple works have focused on resource allocation from an economic perspective, with the goal of maximizing patient benefits and healthcare system efficiency \cite{rw_3_4,rw_3_5}. For instance, Ashana et al \cite{rw_3_6} study the correct resource allocation protocol for hospitals in the United States, revealing these may lead to racial disparities in resource allocation. The authors suggested that more equitable mortality prediction scores are needed which can be partially achieved by computational models rather than post-hoc human decision making. In a similar manner, Farrell et al. \cite{rw_3_7} reviewed the age-based resource allocation during the COVID-19 pandemic, statistically showing the utilized programs are sub-optimal as age does not found to be a good indicator for the resources a patient needs. 

To address resource allocation challenges within hospitals, optimization models have been proposed \cite{rw_3_ai_1}. Lazebnik \cite{teddy_hospitals}, proposed a deep reinforcement learning-based model that uses agent-based simulation with limited historical data to suggest stuff and recourse allocation policies for a wide range of objectives. The model considered factors such as patient demand, resource availability, and treatment priorities, resulting in an optimal allocation of beds, staff, and medical supplies to improve patient care. In a related study, Elitzur et al. \cite{new_rw_7} demonstrated the synergy of predictive analytics methods utilizing machine learning algorithms with optimal pre-test screening protocols. This fusion aims to enhance test efficiency and potentially enable healthcare practitioners to render treatment-related decisions leveraging partial test results without significantly diminishing overall treatment effectiveness. Likewise, Xu et al. \cite{new_rw_8} introduced a model grounded in reinforcement learning to manage an elective surgery backlog post-pandemic disruptions. The model's efficacy was tested using simulated datasets derived from a China-based hospital's elective surgery backlog in the aftermath of the COVID-19 outbreak. Notably, these works consider the patient population to be identical, different only by their clinical needs, assuming static clinical practices over time as well as ignoring entirely possible personalized medicine requirements.  

\subsection{Bladder cancer personalized treatments}
\label{sec:bladder_cancer}
Bladder cancer (BC), a prevalent malignancy, has garnered substantial attention in the realm of personalized medicine due to its clinical heterogeneity and varying treatment responses \cite{agent_spatial_location}. 
A comprehensive review of existing literature reveals diverse efforts aimed at tailoring treatments to individual patients \cite{Precision, Precision1, Precision2}. BC can be classified into non-invasive muscle cancer (NMIBC) or muscle-invasive bladder cancer (MIBC) subtypes depending on genetic background and clinical prognosis. Until now, the gold standard and confirmed diagnosis of BC is cystoscopy, and the main problems of BC are the high recurrence rate and high costs in the clinic \cite{morf}.

For the first group, the standard treatment involves surgical removal of the visible tumor (i.e., transurethral resection) \cite{DeGeorge}. NMIBC tends to progress, so transurethral resection is usually followed by a 6-weekly immunotherapy treatment \cite{Lamm_Morales}. For some patients, this standard treatment is ineffective. For them, multiple clinical options are developed to stop the progress of cancer \cite{stopcancer}, \say{Nadofaragene Firadenovec} gene therapy \cite{genetherapy, firadenovec}, maintenance therapy \cite{maintenance}, and anticancer chemotherapy drugs \cite{Griffiths}. For the same purpose, to stop the BC progress, RNA-binding proteins are being studied, which play a critical and multifaceted role in oncogenesis, and in the prognosis of BC, and, apparently, are most suitable for personalized, say, initial treatment, i.e. changes in the general protocol treatment\cite{related_genes,related_genes1}. In addition, recent molecular and genetic studies have identified new biomarkers and potential therapeutic targets for BC. Indeed, Kiselyov et al. \cite{biomarker1} show the combination of the mathematical methods with molecular and cellular biology insight in the clinical input to receive the individual protocol for every BC patient. Similarly, Blanca et al. \cite{miRNA} reveal the understanding of miRNAs mechanisms and cell distribution provides new opportunities for diagnosis, prognostic, disease monitoring, and personalized therapy of BC patients. In addition, multiple computational and mathematical models investigate the possibility to integrate more advanced personalization systems for BC treatment \cite{oldPaperOde,extend_biology_katya,newPaper}. For instance, Bunimovich-Mendrazitsky et al. \cite{BCG_and_IL} used ordinary differential equations as the guide to find the calculation of optimal treatment protocol. Lazebnik et al.\cite{oldPaper} used partial differential equations to find the optimal treatment protocol considering the location of the tumor and its size. 

For the second group, there is a limited number of variations in treatment types \cite{invasive1}. Multimodal treatment involving radical cystectomy with neoadjuvant chemotherapy offers the best chance for cure, in this case, \cite{ibc_global}. However, the treatment protocol is not personalized and as such (partially) results in a wide range of clinical outcomes. Recently, Su et al.\cite{Precision} proposed new therapies based on deep knowledge of molecular mechanisms of carcinogenesis that have emerged in the clinic, which improved the accuracy of MIBC treatment, and improved prognosis. In a similar manner, Kiselyov et al.\cite{biomarker2} show that a multidisciplinary approach involving simulation, molecular biology,
and clinical science may yield a real opportunity to increase the disease-free and overall survival of patients.

\section{Model Definition}
\label{sec:model}
\begin{figure}[!ht]
    \centering
    \includegraphics[width=0.5\textwidth]{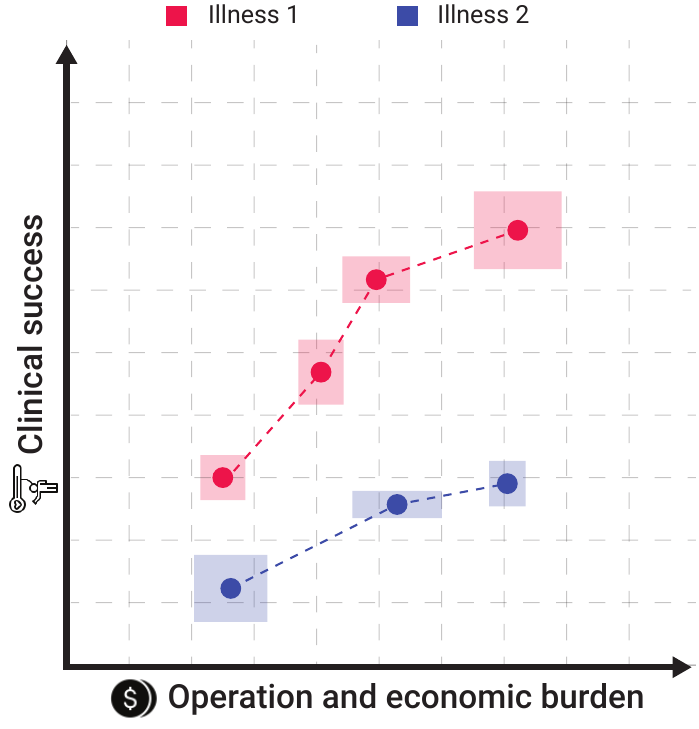}
    \caption{A schematic view of the proposed model's settings. The dots of the same color indicate different levels of personalization for the same treatment. The boxes sounding the dots present the different versatility of this treatment configuration in terms of both the CSR and OEB.}
    \label{fig:model_scheme}
\end{figure}

Intuitively, hospitals are aiming to save the lives of as many of their patients under s strict economic constrain. In this setup, healthcare professionals are able to decide the level of personalization they provide to each patient under some pre-defined set of available treatment configuration for each illness. Each treatment configuration can be associated with two parameters: the clinical success rate (CSR) and the operation and economic burden (OEB). More often than not, as the CSR of a treatment configuration is higher than its alternative its OEB is also higher as well. Moreover, different illnesses has different set of available treatment configurations. Fig. \ref{fig:model_scheme} presents a schematic view of two sets of treatment configurations such that the x-axis indicates the operation and economic burden associated with some treatment configuration while the y-axis indicates the clinical success rate of such treatment. The dots of the same color indicate different levels of personalization for the same treatment. The boxes sounding the dots present the different versatility of this treatment configuration in terms of both the CSR and OEB metrics. 

Formally, the model is focused on the resource allocation of a HSP at some point in time to treat a set of patients. As each patient \(p \in P\) from a population of size \(n \in \mathbb{N}\) is associated with some illness from a finite set, \(I\). For each illness \(i \in I\), there is a finite set of treatments represented by a tuple \((c, o)\) where \(c \in [0, 1]\) is the mean CSR and \(o \in \mathbb{R}^+\) is the mean OEB. Overall, for all illnesses, there is a finite set of size \(z \in \mathbb{N}\) of treatment protocols available to the HSP. The HSP needs to provide each patient treatment from the available set of treatments aiming to increase the CSR of the patient while also satisfying a limited budget constraint \(b \in \mathbb{R}^+\). Therefore, let us define \(x_{i,j}\) a binary integer variable that indicates that the \(i_{th}\) patient obtains the \(j_{th}\) treatment configuration. In addition, if a treatment \(j\) is not relevant to the illness that the \(i_{th}\) patient has, its CSR is set to 0.  Notably, each patient is allowed to obtain only one treatment. Hence, the proposed models take the following form:
\begin{equation}
    \begin{array}{l}
   \max_{x_{i,j}} \sum_{i=1}^n \sum_{j=1}^z c_{j} x_{i,j} \\ \\
   s.t. \\ 
   \sum_{i=1}^n \sum_{j=1}^z x_{i,j} < b, \\
   
   \forall j \in [0, \dots, z]: \sum_{i=1}^n o_j x_{i,j} = 1, \\ 
   
   \forall i \in [0, \dots, n] \wedge j \in [0, \dots, z]: x_{i,j} \leq 1, \\
   
   \forall i \in [0, \dots, n] \wedge j \in [0, \dots, z]: x_{i,j} \geq 0.\\
    
    \end{array}
\end{equation}

\section{In-silico Analysis}
\label{sec:results}
For the purpose of investigating the proposed model in a realistic (yet simplified) scenario, we focused on the bladder cancer types of illnesses. First, we outline the setup of the \textit{in silico} experiment with all the model's parameter values. Then, we present the obtained results and their analysis. 

\subsection{Setup}
In order to investigate the proposed model, one is required to realize a realistic configuration for the model and solve it. Focusing on the latter, as the proposed model takes the form of an integer programming, one can efficiently solve it using the \textit{Simplex} algorithm \cite{simplex}. The result of such computation is a map function between a patient and the treatment configuration. We shall refer to it as the HSP's \textit{policy}. For the purpose of realizing a realistic HSPs' requirement to treat a patient population, we used the data and synthetic data simulator proposed by \cite{teddy_hospitals} which is based on real-world data from four community HSPs that includes the patient population sizes, their illness distribution, obtained treatments, and estimation to the overall OEB. 

Nonetheless, since data about the different treatment configurations for each illness is challenging to obtain for all the treatments a standard HSP is providing, we focused this work on BC diseases. To this end, following Section \ref{sec:bladder_cancer}, we assume two main illnesses: invasive and non-invasive cancer BC. For the invasive and non-invasive BC, there are \(3\) and \(2\) treatment configurations, respectively. Table \ref{table:treatments_bc} presents a summary of the available treatment configuration with their respective CSR and OEB properties. The CSR is represented as the mean \(\pm\) standard deviation as reported by the sources. Notably, as the personalized treatment is not yet clinically validated, these results obtained from \textit{in silico} experiments and will be used lacking any more clinically-established data. The OEB is presented after normalizing for the non-personalized treatment protocol as a baseline. Moreover, as the OEB may change over time, country, and even the HSP itself we computed the OEB according to the operational cost in Israel for 2022 based on the set of services and tests required to provide each treatment configuration. The full description of the OEB computation is provided as supplementary material. 

\begin{table}[!ht]
\centering
\begin{tabular}{ccccc}
\hline \hline
\textbf{Illness}     & \textbf{Treatment} & \textbf{CSR} & \textbf{Normalized OEB} & \textbf{Sources} \\ \hline \hline
\multirow{3}{*}{Non-invasive BC} & Global & \(0.64 \pm 0.08\) & \(1.00\) &  \cite{MoralesEidingerBruce}    \\
   & Initial treatment personalization &  \(0.71 \pm 0.07\) & \(1.07\) & \cite{teddy_cells}     \\
   & During treatment personalization &  \(0.75 \pm 0.04\) & \(1.18\) &   \cite{liza_before_phd}   \\ \hline
\multirow{2}{*}{Invasive BC} & Global &  \(0.32 \pm 0.03\) & \(1.32\) &  \cite{ibc_global}    \\
   & Initial treatment personalization &  \(0.36 \pm 0.03\) & \(1.38\) &  \cite{ibc_second} \\ \hline \hline
\end{tabular}
\caption{A summary of the treatment configurations with their CSR and OEB data used as part of the simulation.}
\label{table:treatments_bc}
\end{table}

In order to measure the level of the policies' personalization, we define the following metric:
\begin{equation}
    \rho := \frac{1}{n}\sum_{i=1}^n \sum_{j=1}^z \frac{c_j - a_j}{b_j - a_j} x_{i,j},
\end{equation}
where \(a_j := \min_j c_j\) and  \(b_j := \max_j c_j\). The motivation for this metric lies in its edge cases. If all patients would get the global treatment, then the personalization value would be 0, according to the proposed metric. Similarly, for the other end of the spectrum, choosing the most personalized treatment protocol for each would result in a value 1.

In addition, one is required to define the budget available for the HSP (\(b\)). Since we know the budget is at least large enough to cover the global treatments of both types of illnesses, then \(b\) is larger than the number of non-invasive BC patients multiplied by the global treatment protocol cost plus the number of invasive BC patients multiplied by the global treatment protocol cost. In addition, it is known that HSPs keep an extra budget for after-treatment care and complications \cite{cost_extra}. We assume, following common practices \cite{budget_up_1,budget_up_2}, that this sum is around 15\% of \(b\), and up to half of it might be utilized for the personalization of treatment protocols. Formally, let us assume \(n_1\) non-invasive BC patients and \(n_2\) invasive BC patients with associated costs \(b_1\) and \(b_2\), respectively. Hence, \(b := (1 + f) \cdot (n_1 b_1 + n_2 b_2)\), where \(f ( = 0.075)\) is the overhead budget provided for the treatment personalization purpose. Moreover, we samples \(n_1 \in [20, 200]\) in a uniform manner. Following historical statistics, \(n_2 \sim 0.08n_1\) \cite{bc_portion}, hence we also samples \(n_2 \in [0.06n_1, 0.1n_1]\) in a uniform manner.  

Based on this configuration, we are interested to answer three main questions:
\begin{itemize}
    \item The relationship between an overhead budget provided for the treatment personalization, \(f\), and the HSP's optimal policy's personalization level, \(\rho\)?
    \item When and how much each personalization treatment protocol is preferred over the others?
    \item How robust are the HSP's policies to the model's parameters (\(n, c, o\))? 
\end{itemize}

\subsection{Results}
In order to solve the proposed model for each instance, we conducted a two-step process. We first constructed the instance using the Python programming language (version 3.8.1) \cite{python}. Afterward, we use the simplex solver created by IBM \cite{ibm} to solve the instance and analyze the results.

\begin{figure}[!ht]
    \centering
    \includegraphics[width=0.99\textwidth]{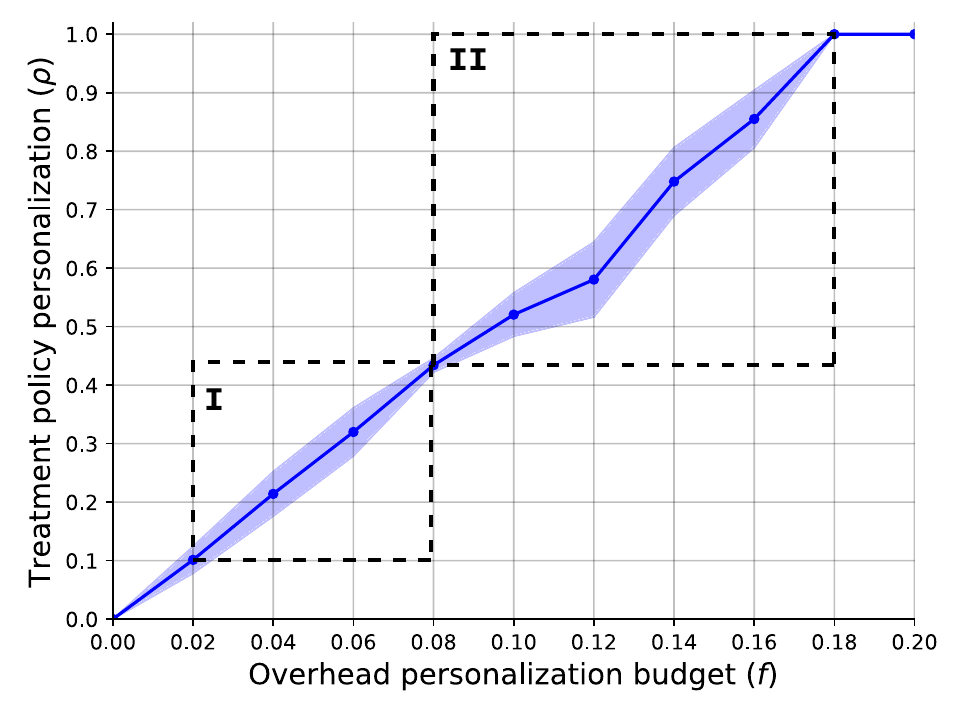}
    \caption{A function of the SP's optimal policy's personalization level (\(\rho\)) with respect to the overhead budget provided for the treatment personalization (\(f\)). The results are shown as the mean \(\pm\) standard deviation of \(n=1000\) repetitions. }
    \label{fig:q1}
\end{figure}

In order to answer the first question, we computed the relationship between the overhead budget provided for the treatment personalization, \(f\), and the HSP's optimal policy's personalization level. Fig. \ref{fig:q1} presents the results of this analysis such that the plot indicates the mean \(\pm\) standard deviation of \(n=1000\) repetitions. The graph shows a linear correlation between the two quantas which is \(\rho = 0.53f - 0.03\), obtained using the least mean square method \cite{lma} with a coefficient of determination of \(R^2 = 0.92\). One can notice two sharp increases in the policy's personalization levels' standard deviation, as marked by \(I\) and \(II\). The first increase and then decrease can be associated with the fact that \(f=0.08\) allows to provide all the non-invasive BC patients the initial treatment personalization which is (\(\sim 0.4\)) of the personalization level for this illness due to the last treatment in the category. By the same token, the second increase starts right after and returns to zero when \(f = 0.18\) allowing to provide all patients the most personalized treatment for each illness. 

\begin{figure}[!ht]
    \centering
    \includegraphics[width=0.99\textwidth]{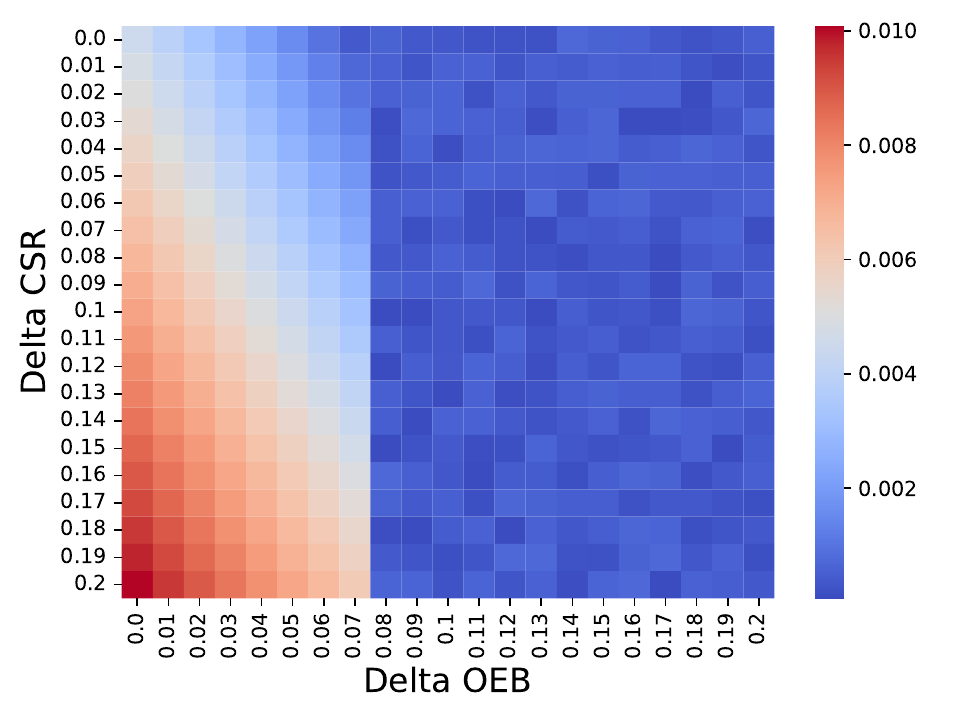}
    \caption{The normalized number of choosing each treatment protocol based on its delta OEB and CSR compared to the global treatment (e.g., the non-personalized treatment). The results are shown as the mean of \(n=1000\) repetitions.}
    \label{fig:q2}
\end{figure}

As a means to answer the second question, we ignore the possible treatment protocols for BC (as shown in Table \ref{table:treatments_bc}) and generate abstract treatment protocols as follows. A treatment \(a\) is set to be more personalized than treatment \(b\) if and only if the latter's OEB and mean CSR are smaller compared to these of \(a\). For each iteration of the model, we generate between one and four personalized treatments such that the most personalized treatment has an OEB of up to 25\% higher than the global treatment and 20\% improvement in the CSR. In addition, unlike the previous analysis, we set \(f\) to be \(0.075\). Using this configuration, we store the delta OEB and delta CSR between the chosen personalized treatment and the global one for each patient, computing their chosen distribution over \(n = 1000\). Fig. \ref{fig:q2} shows a heatmap that indicates the normalized number of choices with democratization of one percent. It is easy to notice the phase transition for delta OEB of 0.07 to 0.08. This transition can be explained by the fact that \(f=0.075\) so only part of the population can obtain it and if a less \say{expensive} alternative is present, the model would prefer it to save more people, on average. Moreover, using SciMed \cite{scimed} a symbolic regression tool that utilizes a genetic-algorithm-based approach \cite{ga_intro} to search for an analytical function that best fits data, we computed that for OEB of 0.07 or less, the function
\begin{equation}
    count = 0.32 - 1.95 \Delta OEB - 0.09 (\Delta OEB)^2 + 4.07 \Delta CSR + 0.56 \Delta OEB \cdot \Delta CSR, 
    \label{eq:fit_q2}
\end{equation} best explains the dynamics with a coefficient of determination of \(R^2 = 0.88\). 

\begin{figure}[!ht]
\centering

\begin{subfigure}{.49\linewidth}
    \centering
    \includegraphics[width=0.99\linewidth]{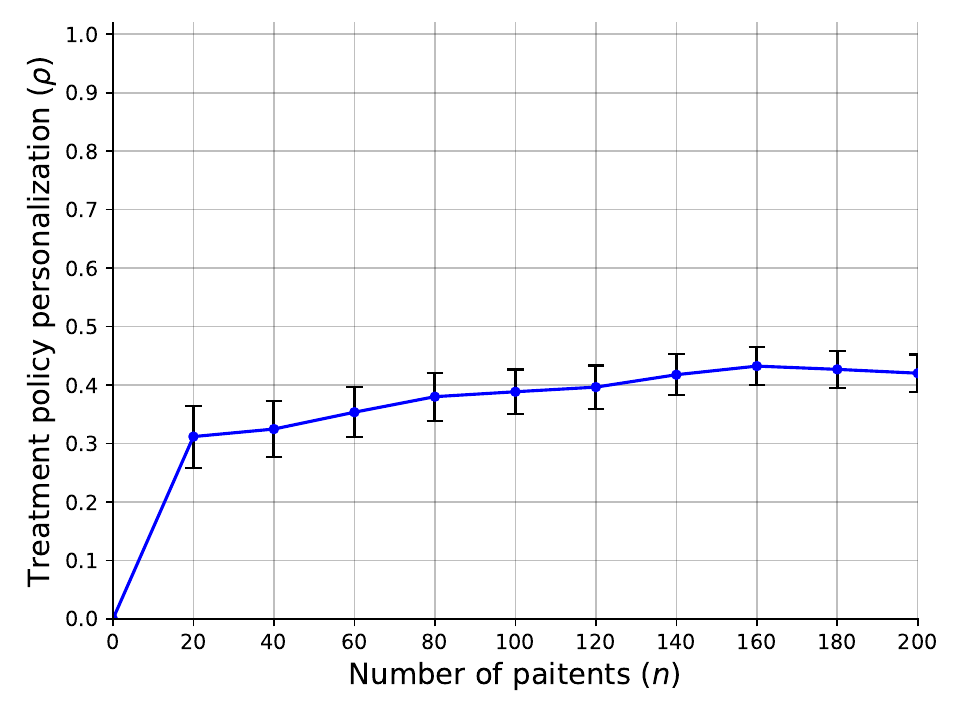}
    \caption{Number of patients.}
    \label{fig:q3_n}
\end{subfigure}

\begin{subfigure}{.49\linewidth}
    \centering
    \includegraphics[width=0.99\linewidth]{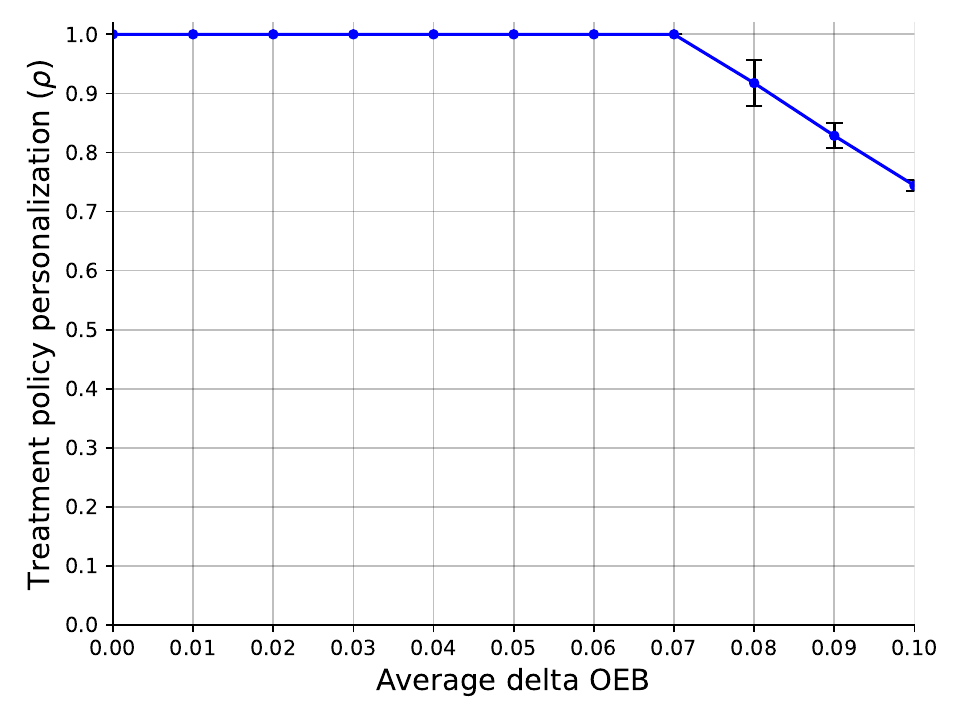}
    \caption{Average delta OEB.}
    \label{fig:q3_c}
\end{subfigure}
\begin{subfigure}{.49\linewidth}
    \centering
    \includegraphics[width=0.99\linewidth]{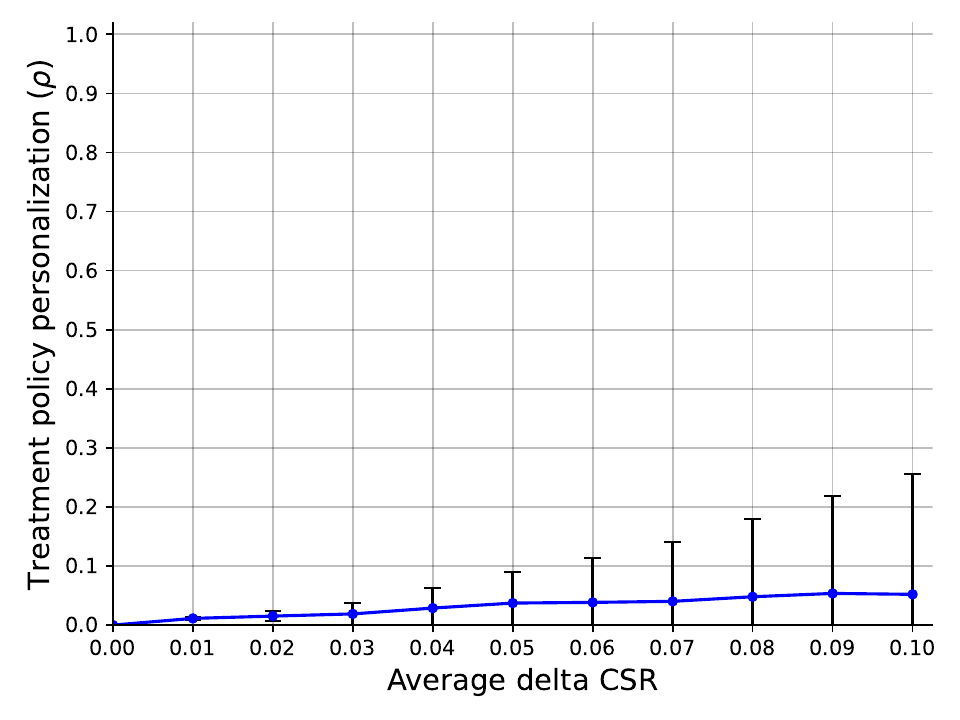}
    \caption{Average delta CSR.}
    \label{fig:q3_o}
\end{subfigure}

\caption{A sensitivity analysis of the treatment policy personalization level \((\rho)\) with respect to the model's parameters. The results are shown as an mean \(\pm\) standard deviation of \(n=1000\) repetitions.}
\label{fig:q3}
\end{figure}

Regarding the third question, we computed the sensitivity analysis of the treatment policy personalization level (\(\rho\)) with respect to the number of patients, average delta OEB, and average delta CSR. Fig. \ref{fig:q3} presents the results as the mean \(\pm\) standard deviation of \(n = 1000\) repetitions. In particular, Fig. \ref{fig:q3_n} shows that a larger number of patients allows for a slight increase in the overall treatment policy personalization level which keeps a similar level of stability (as indicated by the error bars). In addition, Fig. \ref{fig:q3_c} shows that when the OEB is smaller than \(f = 0.075\) all the population can obtain the personalized treatment results in \(\rho = 1\). However, a linear decrease is present once the average OEB crosses the \(f = 0.075\) threshold. Finally, Fig. \ref{fig:q3_o} shows that as the CSR increase, the treatment policy personalization level also increases. Interestingly, the standard deviation is also increasing and even quicker than the \(\rho\) itself.

\section{Discussion and Conclusion}
\label{sec:discussion}
In this study, we propose a mathematical model to investigate the usage of a spectrum of personalized treatments by a healthcare service provider (HSP) under an operation and economic burden (OEB) limitation as the HSP has a limited budget. The model is formalized using an integer programming task where the optimal allocation of treatment to a patient is computed. Using the proposed model and (psudo-)realistic data configuration, we investigated the proposed model and its implementation for hospitals in the context of BC treatment. 

Our findings provide valuable insights into the intricate relationship between treatment personalization and economic considerations. Fig. \ref{fig:q1} depicts the clear linear correlation between the overhead budget allocated for treatment personalization and the level of personalization within the HSP's policy. The gradual increase in budget allocation results in a proportional rise in the availability of personalized treatment options for patients. To be specific, the illness with the larger patient population dominates the other illness from the HSP's point of view as the overhead cost associated with these patients is less dominant in the improvement of the HSP's objective. In addition, as budgets are somewhere between the cost of two personalization treatments, a more complex resource allocation task is raised which causes the large differences in different instances of the model. Hence, while larger budgets shown a linear increase in the personalization level and therefore CSR, one should take into consideration these dynamics in allocating the budget (\(f\)). 

Furthermore, Fig. \ref{fig:q2} elucidates the dynamic interplay between treatments increases in the clinical success rate and their operational and economic burden. Namely, Eq. (\ref{eq:fit_q2}) shows that a second-order polynomial decrease in the HSP's policy personalization level as the OEB of these treatments increases while an increase in the CSR results in an increase in the HSP's policy personalization level. Surprisingly, a linear multiplication between the two also politically correlated with the HSP's policy personalization level. This outcome reveals the quite straightforward result that higher CSR and lower OEB would allow HSP's to perform more personalized treatment. However, the interesting outcome is that one can focus on improving the treatment for more or less the same price and gain almost twice as much result for small increases, as indicated by the coefficients in the formula. 

Compellingly, Fig. \ref{fig:q3_n} reveals that an increased number of patients allows HSPs to utilize, on average, more personalized treatment policies. This outcome well-aligns with known goods production theory and practice \cite{end_1,end_2,end_3}. In a complementary manner, Fig. \ref{fig:q3_c} shows that in the case that the overhead personalization budget is larger than the personalization cost, all patients would get it. In scenarios where this is not the case, the portion of the two values determines the number of patients who would get the personalized treatment. While this outcome is somewhat trivial, it is a good sanity check for the proposed model's ability to capture the discussed clinical-economic dynamic. On the other hand, as the CSR of treatment increases, it is more likely to \say{justify} its CSR and be included as part of the HSP's policy, on average. That said, as the CSR of each such personalized treatment plays a central role in the decision, we can observe an increase in the instability of the treatment policy personalization, as indicated by the increase in the error bars. 

Taken jointly, our results reveal a promising future for personalized medicine. Due to their clear clinical advantage, policymakers in HSPs can (and should) allocate a budget, which itself would be available as a result of decreasing the amount of after-treatment care-related clinical services, for personalized treatments. Moreover, we show that one can either improve the cost of personalized medicine or its clinical performance to allow HSPs to utilize them more. To this end, when considering small improvement steps, the improvement in clinical performance outperforms the improvement in cost reduction which is known to be relatively quicker to achieve \cite{end_4,end_5,end_6}. 

The proposed work is not without limitations which provide a venue for future work. First, the proposed model is focused on a single point in time, simplifying the temporal complexity of the task as patients' populations and treatments change \cite{intro_10,intro_9,intro_11}. Second, the HSP's budget is not the only source of support for patients' treatment economic burden as patients may acquire services out of pocket or through insurance. Taking these external funding resources in a future model would make it more realistic. Third, giving more data about the possible treatment configurations and their OED would allow us to investigate the proposed model on a larger and more realistic scale, obtaining finer results. Finally, patients are assumed to be homogeneous, differing only by their illness. However, this is only an approximation that should be relaxed in future work. 

\section*{Declarations}
\subsection*{Funding}
This research has been supported by a Grant from Ariel University. 

\subsection*{Conflicts of interest/Competing interests}
None.

\subsection*{Data availability}
All the data used in this study is available online from the referenced sources. 

\subsection*{Acknowledgement}
Elizaveta Savchenko wishes to thank Ariel University's financial support during this research. The authors wish to thank Teddy Lazebnik for helping with the modeling and analysis presented in this work.

\subsection*{Author Contribution}
Elizaveta Savchenko: Conceptualization, Methodology, Software, Formal analysis, Investigation, Visualization, Writing - Original Draft. \\  
Svetlana Bunimovich-Mendrazitsky: Validation, Supervision, Writing - Review \& Editing. \\ 
 
\bibliography{biblio}

\begin{thebibliography}{10}

\bibitem{start_1}
J.~E. Schneider, T.~R. Miller, R.~L. Ohsfeldt, M.~A. Morrisey, B.~A. Zelner,
  and P.~Li.
\newblock The economics of specialty hospitals.
\newblock {\em Medical Care Research and Review}, 65(5):531--553, 2008.

\bibitem{start_2}
M.~W. Reder.
\newblock Some problems in the economics of hospitals.
\newblock {\em The American Economic Review}, 55(1):472–480, 1965.

\bibitem{reviewer_ask}
I.~Ahn and J.~Park.
\newblock Drug scheduling of cancer chemotherapy based on natural actor-critic
  approach.
\newblock {\em Biosystems}, 106(2):121--129, 2011.

\bibitem{teddy_bcg_suger}
T.~Lazebnik, S.~Bunimovich-Mendrazitsky, and A.~Kiselyov.
\newblock Mathematical model for bcg-based treatment of type 1 diabetes.
\newblock {\em Physica A: Statistical Mechanics and its Applications},
  622:128891, 2023.

\bibitem{liza_phd_1}
E.~Savchenko, A.~Rosenfeld, and S.~Bunimovich-Mendrazitsky.
\newblock Mathematical modeling of bcg-based bladder cancer treatment using
  socio-demographics.
\newblock {\em arXiv}, 2023.

\bibitem{teddy_cells}
T.~Lazebnik.
\newblock Cell-level spatio-temporal model for a bacillus
  calmette–guérin-based immunotherapy treatment protocol of superficial
  bladder cancer.
\newblock {\em Cells}, 15(11):2372, 2022.

\bibitem{journal_2}
R.~Elitzur, D.~Krass, and E.~Zimlichman.
\newblock Machine learning for optimal test admission in the presence of
  resource constraints.
\newblock {\em Health Care Management Science}, 2023.

\bibitem{intro_1}
V.~Khashayar, R.~Jason, R.~Linda, and F.~Samir.
\newblock Optimizing physician staffing and resource allocation: Sine-wave
  variation in hourly trauma admission.
\newblock {\em The Journal of Trauma: Injury, Infection, and Critical Care},
  62(3):610--614, 2007.

\bibitem{intro_2}
A.~D. Asante and A.~B. Zwi.
\newblock Factors influencing resource allocation decisions and equity in the
  health system of ghana.
\newblock {\em Public Health}, 123(5):371--377, 2009.

\bibitem{paitent_happy}
H.~Vuori.
\newblock {Patient Satisfaction - Does It Matter?}
\newblock {\em International Journal for Quality in Health Care},
  3(3):183--189, 1991.

\bibitem{per_1}
N.~J. Schork.
\newblock {\em Artificial Intelligence and Personalized Medicine}, pages
  265--283.
\newblock 2019.

\bibitem{per_2}
K.~K Jain.
\newblock Personalized medicine.
\newblock {\em Current opinion in molecular therapeutics}, 4(6), 2002.

\bibitem{per_3}
I.~S. Chan and G.~S. Ginsburg.
\newblock Personalized medicine: Progress and promise.
\newblock {\em Annual Review of Genomics and Human Genetics}, 12(1):217--244,
  2011.

\bibitem{per_happy_1}
B.~Prainsack.
\newblock Personhood and solidarity: what kind of personalized medicine do we
  want?
\newblock {\em Personalized Medicine}, 11(7):651--657, 2014.

\bibitem{per_happy_2}
B.~Prainsack.
\newblock The “we” in the “me”: Solidarity and health care in the era
  of personalized medicine.
\newblock {\em Science, Technology, \& Human Values}, 43(1):21--44, 2018.

\bibitem{per_method_1}
A.~Ziegler, A.~Koch, K.~Krockenberger, and A.~Grobhennig.
\newblock Personalized medicine using dna biomarkers: a review.
\newblock {\em Human Genetics}, 131:1627–1638, 2012.

\bibitem{per_method_2}
T.~Lazebnik, Z.~Bahouth, S.~Bunimovich-Mendrazitsky, and S.~Halachmi.
\newblock Predicting acute kidney injury following open partial nephrectomy
  treatment using sat-pruned explainable machine learning model.
\newblock {\em BMC Medical Informatics and Decision Making}, 22:133, 2022.

\bibitem{per_method_3}
M.~Whirl-Carrillo, E.~M. McDonagh, J.~M. Hebert, L.~Gong, K.~Sangkuhl, C.~F.
  Thorn, R.~B. Altman, and T.~E. Klein.
\newblock Pharmacogenomics knowledge for personalized medicine.
\newblock {\em Clinical Pharmacology \& Therapeutics}, 92(4):414--417, 2012.

\bibitem{teddy_hospitals}
T.~Lazebnik.
\newblock Data-driven hospitals staff and resources allocation using
  agent-based simulation and deep reinforcement learning.
\newblock {\em Engineering Applications of Artificial Intelligence},
  126:106783, 2023.

\bibitem{ml_example_1}
A.~Zlotnik, A.~Gallardo-Antolin, M.~C. Alfaro, M.~C.~R. Perez, and J.~M.~M.
  Martinez.
\newblock Emergency department visit forecasting and dynamic nursing staff
  allocation using machine learning techniques with readily available
  open-source software.
\newblock {\em Computers, Informatics, Nursing}, 33(8):368--377, 2015.

\bibitem{ml_example_2}
B.~Lehaney and V.~Hlupic.
\newblock Simulation modelling for resource allocation and planning in the
  health sector.
\newblock {\em Journal of the Royal Society of Health}, 115(6):382--385, 1995.

\bibitem{ml_example_3}
M.~Liu and D.~Zhang.
\newblock A dynamic logistics model for medical resources allocation in an
  epidemic control with demand forecast updating.
\newblock {\em Journal of the Operational Research Society}, 67(6):841--852,
  2016.

\bibitem{eco_good_1}
A.~Alexi, T.~Lazebnik, and L.~Shami.
\newblock Microfounded tax revenue forecast model with heterogeneous population
  and genetic algorithm approach.
\newblock {\em Computational Economy}, 2023.

\bibitem{eco_good_2}
Warwick McKibbin and David Vines.
\newblock Global macroeconomic cooperation in response to the covid-19
  pandemic: a roadmap for the g20 and the imf.
\newblock {\em Oxford Review of Economic Policy}, 36(Supplement\_1):S297--S337,
  2020.

\bibitem{eco_good_3}
H.~Rutter, N.~Savona, K.~Glonti, J.~Bibby, S.~Cummins, D.~T. Finegood,
  F.~Greaves, L.~Harper, P.~Hawe, L.~Moore, M.~Petticrew, E.~Rehfuess,
  A.~Shiell, J.~Thomas, and M.~White.
\newblock The need for a complex systems model of evidence for public health.
\newblock {\em The Lancet}, 390:2602--2604, 2017.

\bibitem{eco_good_4}
S.~Zheng, A.~Trott, S.~Srinivasa, D.~C. Parkes, and R.~Socher.
\newblock The ai economist: Optimal economic policy design via two-level deep
  reinforcement learning.
\newblock {\em arXiv}, 2021.

\bibitem{intro_7}
J.~E. Harris.
\newblock The internal organization of hospitals: Some economic implications.
\newblock {\em The Bell Journal of Economics}, 8(2):467--482, 1977.

\bibitem{intro_8}
S.~Talati, P.~Bhatia, A.~Kumar, A.~K. Gupta, and C.~D. Ojha.
\newblock Strategic planning and designing of a hospital disaster manual in a
  tertiary care, teaching, research and referral institute in india.
\newblock {\em World J Emerg Med}, 5(1):35--41, 2014.

\bibitem{model_intro_4}
J.~J. Patard, F.~Saint, F.~Velotti, C.~C. Abbou, and D.~K. Chopin.
\newblock Immune response following intravesical bacillus calmette–guérin
  instillations in superficial bladder cancer: a review.
\newblock {\em Urological Research}, 26(3):155--159, 1998.

\bibitem{cancer_intro_3}
J.~L. McCall, M.~R. Cox, and D.~A. Wattchow.
\newblock Analysis of local recurrence rates after surgery alone for rectal
  cancer.
\newblock {\em Int J Colorect Dis}, 10:126--132, 1995.

\bibitem{intro_6}
M.~Moleman, T.~Zuiderent-Jerak, M.~Lageweg, G.~L. van~den Braak, and T.~J.
  Schuitmaker-Warnaar.
\newblock Doctors as resource stewards? translating high-value, cost-conscious
  care to the consulting room.
\newblock {\em Health Care Analysis}, 30:215–239, 2022.

\bibitem{rw_2_1}
D.~Cirillo and A.~Valencia.
\newblock Big data analytics for personalized medicine.
\newblock {\em Current Opinion in Biotechnology}, 58:161--167, 2019.

\bibitem{rw_2_2}
V.~Gambardella, N.~Tarazona, J.M. Cejalvo, P.~Lombardi, M.~Huerta, S.~Roselló,
  T.~Fleitas, D.~Roda, and A.~Cervantes.
\newblock Personalized medicine: Recent progress in cancer therapy.
\newblock {\em Cancers}, 12:1009, 2020.

\bibitem{rw_2_3}
S.~Morand, M.~Devanaboyina, H.~Staats, L.~Stanbery, and J.~Nemunaitis.
\newblock Ovarian cancer immunotherapy and personalized medicine.
\newblock {\em International Journal of Molecular Sciences}, 22(12), 2021.

\bibitem{rw_2_4}
A.~Blasiak, J.~Khong, and T.~Kee.
\newblock Curate.ai: Optimizing personalized medicine with artificial
  intelligence.
\newblock {\em SLAS TECHNOLOGY: Translating Life Sciences Innovation}, 25(2),
  2020.

\bibitem{rw_2_5}
S.~Morganti, P.~Tarantino, E.~Ferraro, P.~D'Amico, B.~A. Duso, and
  G.~Curigliano.
\newblock {\em Next Generation Sequencing (NGS): A Revolutionary Technology in
  Pharmacogenomics and Personalized Medicine in Cancer}, pages 9--30.
\newblock 2019.

\bibitem{rw_2_6}
M.~Ce, G.~Irmici, C.~Foschini, G.~M. Danesini, L.~V. Falsitta, M.~L. Serio,
  A.~Fontana, C.~Martinenghi, G.~Oliva, and M.~Cellina.
\newblock Artificial intelligence in brain tumor imaging: A step toward
  personalized medicine.
\newblock {\em Current Oncology}, 30(3):2673--2701, 2023.

\bibitem{rw_2_7}
H.~Mzoughi, I.~Njeh, A.~Wali, M.B. Slima, A.~BenHamida, C.~Mhiri, and K.B.
  Mahfoudhe.
\newblock Deep multi-scale 3d convolutional neural network (cnn) for mri
  gliomas brain tumor classification.
\newblock {\em J. Digit. Imaging}, 33:903–915, 2020.

\bibitem{rw_2_8}
J.~J. Van~Netten, J.~Woodburn, and S.~A. Bus.
\newblock The future for diabetic foot ulcer prevention: A paradigm shift from
  stratified healthcare towards personalized medicine.
\newblock {\em Diabetes/Metabolism Research and Reviews}, 36(S1):e3234, 2020.

\bibitem{liza_before_phd}
A.~Yaniv-Rosenfeld, E.~Savchenko, A.~Rosenfeld, and T.~Lazebnik.
\newblock Scheduling bcg and il-2 injections for bladder cancer immunotherapy
  treatment.
\newblock {\em Mathematics}, 11(6):1192, 2023.

\bibitem{amit_new}
add later.
\newblock add later.
\newblock {\em add later}, 2023.

\bibitem{rw_3_1}
K.~F. Boreskie, P.~E. Boreskie, and D.~Melady.
\newblock Age is just a number – and so is frailty: Strategies to inform
  resource allocation during the covid-19 pandemic.
\newblock {\em Canadian Journal of Emergency Medicine}, 22(4):411–413, 2020.

\bibitem{rw_3_2}
E.~L. {Daugherty Biddison}, R.~Faden, H.~S. Gwon, D.~P. Mareiniss, A.~C.
  Regenberg, M.~Schoch-Spana, J.~Schwartz, and E.~S. Toner.
\newblock Too many patients…a framework to guide statewide allocation of
  scarce mechanical ventilation during disasters.
\newblock {\em Chest}, 155(4):848--854, 2019.

\bibitem{rw_3_3}
J.~T. J.~M. van Dijck, M.~D. Dijkman, R.~H. Ophuis, G.~C.~W. de~Ruiter, W.~C.
  Peul, and S.~Polinder.
\newblock In-hospital costs after severe traumatic brain injury: A systematic
  review and quality assessment.
\newblock {\em PLOS ONE}, 14(5):1--21, 2019.

\bibitem{rw_3_4}
G.~Melman, A.~Parlikad, and E.~Cameron.
\newblock Balancing scarce hospital resources during the covid-19 pandemic
  using discrete-event simulation.
\newblock {\em Health Care Manag Sci}, 24:356–374, 2021.

\bibitem{rw_3_5}
M.~Ordu, E.~Demir, C.~Tofallis, and M.~M. Gunal.
\newblock A novel healthcare resource allocation decision support tool: A
  forecasting-simulation-optimization approach.
\newblock {\em Journal of the Operational Research Society}, 72(3):485--500,
  2021.

\bibitem{rw_3_6}
D.~C. Ashana, G.~L. Anesi, V.~X. Liu, G.~J. Escobar, C.~Chesley, N.~D. Eneanya,
  G.~E. Weissman, W.~D. Miller, M.~O. Harhay, and S.~D. Halpern.
\newblock Equitably allocating resources during crises: Racial differences in
  mortality prediction models.
\newblock {\em American Journal of Respiratory and Critical Care Medicine},
  204(2), 2021.

\bibitem{rw_3_7}
T.~W. Farrell, L.~E. Ferrante, T.~Brown, L.~Francis, E.~Widera, R.~Rhodes,
  T.~Rosen, U.~Hwang, L.~J. Witt, N.~Thothala, S.~W. Liu, C.~A. Vitale, U.~K.
  Braun, C.~Stephens, and D.~Saliba.
\newblock Ags position statement: Resource allocation strategies and
  age-related considerations in the covid-19 era and beyond.
\newblock {\em Journal of the American Geriatrics Society}, 68(6):1136--1142,
  2020.

\bibitem{rw_3_ai_1}
A.~Kirubarajan, A.~Taher, S.~Khan, and S.~Masood.
\newblock Artificial intelligence in emergency medicine: A scoping review.
\newblock {\em Journal of the American College of Emergency Physicians Open},
  1(6):1691--1702, 2020.

\bibitem{new_rw_7}
R.~Elitzur, D.~Krass, and E.~Zimlichman.
\newblock Machine learning for optimal test admission in the presence of
  resource constraints.
\newblock {\em Health Care Manag}, 2023.

\bibitem{new_rw_8}
H.~Xu, Y.~Fang, C-A. Chou, and L.~Luo.
\newblock A reinforcement learning-based optimal control approach for managing
  an elective surgery backlog after pandemic disruption.
\newblock {\em Health Care Manag}, 2023.

\bibitem{agent_spatial_location}
S.~Bunimovich-Mendrazitsky, V.~Pisarev, and E.~Kashdan.
\newblock Modeling and simulation of a low-grade urinary bladder carcinoma.
\newblock {\em Computers in Biology and Medicine}, 2014.

\bibitem{Precision}
H.~Su, H.~Jiang, T.~Tao, X.~Kang, X.~Zhang, D.~Kang, S.~Li, C.~Li, H.~Wang,
  Z.~Yang, J.~Zhang, and C.~Li.
\newblock Hope and challenge: Precision medicine in bladder cancer.
\newblock {\em Cancer Med.}, 8:1806--1816, 2019.

\bibitem{Precision1}
B.J. Guercio, G.~Iyer, and J.E. Rosenberg.
\newblock Developing precision medicine for bladder cancer.
\newblock {\em Hematol Oncol Clin North Am.}, 35:633--653, 2021.

\bibitem{Precision2}
S.K. Mohanty, A.~Lobo, S.K. Mishra, and L.~Cheng.
\newblock Precision medicine in bladder cancer: Present challenges and future
  directions.
\newblock {\em J Pers Med.}, 13:756, 2023.

\bibitem{morf}
B.~Tuna, K.~Yörükoglu, E.~Düzcan, and et~al.
\newblock Histologic grading of urothelial papillary neoplasms: impact of
  combined grading (two-numbered grading system) on reproducibility.
\newblock {\em Virchows Arch}, 458:659–664, 2011.

\bibitem{DeGeorge}
K.C. DeGeorge, H.R. Holt, and S.C. Hodges.
\newblock Bladder cancer: Diagnosis and treatment.
\newblock {\em Am Fam Physician}, 15(96):507--514, 2017.

\bibitem{Lamm_Morales}
D.L. Lamm and A.~Morales.
\newblock A bcg success story: From prevention of tuberculosis to optimal
  bladder cancer treatment.
\newblock {\em Vaccine}, 8(39):7308--7318, 2021.

\bibitem{stopcancer}
M.S. Cookson, H.W. Herr, Z.F. Zhang, and et~al.
\newblock The treated natural history of high-risk superficial bladder cancer:
  15-year outcome.
\newblock {\em Lancet Oncol}, 158:62--67, 1997.

\bibitem{genetherapy}
S.A. Boorjian, M.~Alemozaffar, B.R. Konety, N.D. Shore, and L.G. et~al.
  Gomella.
\newblock Intravesical nadofaragene firadenovec gene therapy for
  bcg-unresponsive non-muscle-invasive bladder cancer: a single-arm,
  open-label, repeat-dose clinical trial.
\newblock {\em Lancet Oncol}, 22:107--117, 2021.

\bibitem{firadenovec}
G.S. Kulkarni.
\newblock Nadofaragene firadenovec: a new gold standard for bcg-unresponsive
  bladder cancer?
\newblock {\em Lancet Oncol.}, 22:8--9, 2021.

\bibitem{maintenance}
J.~Oddens, M.~Brausi, R.~Sylvester, A.~Bono, C.~van~de Beek, G.~van Andel,
  P.~Gontero, W.~Hoeltl, L.~Turkeri, S.~Marreaud, S.~Collette, and
  W.~Oosterlinck.
\newblock Final results of an eortc-gu cancers group randomized study of
  maintenance bacillus calmette-guérin in intermediate- and high-risk ta, t1
  papillary carcinoma of the urinary bladder: one-third dose versus full dose
  and 1 year versus 3 years of maintenance.
\newblock {\em Eur Urol}, 63:462--472, 2013.

\bibitem{Griffiths}
T.R. Griffiths.
\newblock Current perspectives in bladder cancer management.
\newblock {\em International journal of clinical practice}, 67(5):435--448,
  2013.

\bibitem{related_genes}
C.~Peng, S.~Guo, Z.~Yang, X.~Li, Q.~Su, and W.~Mo.
\newblock A prognostic model for bladder cancer based on cytoskeleton-related
  genes.
\newblock {\em Medicine}, 102, 2023.

\bibitem{related_genes1}
F.~Chen, Q.~Wang, and Y.~Zhou.
\newblock The construction and validation of an rna binding protein-related
  prognostic model for bladder cancer.
\newblock {\em BMC Cancer}, 21:244, 2021.

\bibitem{biomarker1}
Startsev~V. Kiselyov~A, Bunimovich-Mendrazitsky~S.
\newblock Treatment of non-muscle invasive bladder cancer with bacillus
  calmette-guerin (bcg): Biological markers and simulation studies.
\newblock {\em BBA Clin}, 4:27--347, 2015.

\bibitem{miRNA}
A.~Blanca, L.~Cheng, R.~Montironi, and H.~et~al. Moch.
\newblock Mirna expression in bladder cancer and their potential role in
  clinical practice.
\newblock {\em Current Drug Metabolism}, 8:712--722, 2017.

\bibitem{oldPaperOde}
L.~Shaikhet and S.~Bunimovich-Mendrazitsky.
\newblock Stability analysis of delayed immune response bcg infection in
  bladder cancer treatment model by stochastic perturbations.
\newblock {\em Computational and mathematical methods in medicine}, 2018.

\bibitem{extend_biology_katya}
E.~Guzev, S.~Halachmi, and S.~Bunimovich-Mendrazitsky.
\newblock Additional extension of the mathematical model for {BCG}
  immunotherapy of bladder cancer and its validation by auxiliary tools.
\newblock {\em International Journal of Nonlinear Sciences and Numerical
  Simulation}, 20:675--689, 2019.

\bibitem{newPaper}
T.~Lazebnik, S.~Bunimovich-Mendrazitsky, and N.~Haroni.
\newblock {PDE} based geometry model for {BCG} immunotherapy of bladder cancer.
\newblock {\em Biosystems}, 2021.

\bibitem{BCG_and_IL}
S.~Bunimovich-Mendrazitsky and Shaikhet L.
\newblock Stability analysis of delayed tumor-antigen-activated immune response
  in combined bcg and il-2immunotherapy of bladder cancer.
\newblock {\em Processes}, 8:1564, 2020.

\bibitem{oldPaper}
T.~Lazebnik, S.~Yanetz, S.~Bunimovich-Mendrazitsky, and N.~Haroni.
\newblock Treatment of bladder cancer using bcg immunotherapy: Pde modeling.
\newblock {\em Partial Differential Equations}, 2020.

\bibitem{invasive1}
M.~Yin, M.~Joshi, and R.P. et~al. Meijer.
\newblock Neoadjuvant chemotherapy for muscle-invasive bladder cancer: A
  systematic review and two-step meta-analysis.
\newblock {\em Oncologist}, 21:708--15, 2016.

\bibitem{ibc_global}
A.~M. Kamat, N.~M. Hahn, J.~A. Efstanthiou, S.~P. Lerner, P-U. Malmstrom,
  W.~Chai, C.~C. Guo, Y.~Lotan, and W.~Kassouf.
\newblock Bladder cancer.
\newblock {\em Lancet}, 388:2796--2810, 2016.

\bibitem{biomarker2}
Startsev~V. Kiselyov~A, Bunimovich-Mendrazitsky~S.
\newblock Key signaling pathways in the muscle‐invasive bladder carcinoma:
  clinical markers for disease modeling and optimized treatment.
\newblock {\em International Journal of Cancer}, 138:2562--2569, 2016.

\bibitem{simplex}
H.~Konno, Y.~Yajima, and T.~Matsui.
\newblock Parametric simplex algorithms for solving a special class of
  nonconvex minimization problems.
\newblock {\em Journal of Global Optimization}, 1:65--81, 1991.

\bibitem{MoralesEidingerBruce}
A.~Morales, D.~Eidinger, and A.W. Bruce.
\newblock {I}ntracavity {B}acillus {C}almette-{G}u\'{e}rin in the treatment of
  superficial bladder tumors.
\newblock {\em J. Urol}, 116:180--183, 1976.

\bibitem{ibc_second}
T.~Powles, M.~Kockx, A.~Rodriguez-Vida, I.~Duran, M.~S. Crabb, S. J. ad Van
  Der~Heijden, B.~Szabados, A.~F. Pous, G.~Gravis, U.~A. Herranz, A.~Protheroe,
  A.~Ravaud, D.~Maillet, M.~J. Mendez, C.~Suarez, M.~Linch, A.~Prendergast,
  P-J. van Dam, D.~Stanoeva, S.~Daelemans, S.~Mariathasan, J.~S. Tae, K.~Mousa,
  R.~Banchereau, and D.~Castellano.
\newblock Clinical efficacy and biomarker analysis of neoadjuvant atezolizumab
  in operable urothelial carcinoma in the abacus trial.
\newblock {\em Nature Medicine}, 25:1706–1714, 2019.

\bibitem{cost_extra}
E.~Lettieri.
\newblock Uncertainty inclusion in budgeting technology adoption at a hospital
  level: Evidence from a multiple case study.
\newblock {\em Health Policy}, 93(2):128--136, 2009.

\bibitem{budget_up_1}
I.~Lapsley.
\newblock The accounting–clinical interface—implementing budgets for
  hospital doctors.
\newblock {\em Abacus}, 37(1):79--109, 2001.

\bibitem{budget_up_2}
R.~Balakrishnan, N.~S. Soderstrom, and T.~D. West.
\newblock {Spending Patterns with Lapsing Budgets: Evidence from U.S. Army
  Hospitals}.
\newblock {\em Journal of Management Accounting Research}, 19(1):1--23, 2007.

\bibitem{bc_portion}
D.~M. Parkin.
\newblock The global burden of urinary bladder cancer.
\newblock {\em Scandinavian Journal of Urology and Nephrology},
  42(sup218):12--20, 2008.

\bibitem{python}
K.~R. Srinath.
\newblock Python – the fastest growing programming language.
\newblock {\em International Research Journal of Engineering and Technology},
  4(12), 2017.

\bibitem{ibm}
D.~G. Wilson and B.~D. Rudin.
\newblock Introduction to the ibm optimization subroutine library.
\newblock {\em IBM Systems Journal}, 31(1):4--10, 1992.

\bibitem{lma}
M.~K. Transtrum and J.~P. Sethna.
\newblock Improvements to the levenberg-marquardt algorithm for nonlinear
  least-squares minimization.
\newblock {\em arXiv}, 2012.

\bibitem{scimed}
L.~S. Keren, A.~Liberzon, and T.~Lazebnik.
\newblock A computational framework for physics-informed symbolic regression
  with straightforward integration of domain knowledge.
\newblock {\em Scientific Reports}, 13:1249, 2023.

\bibitem{ga_intro}
J.~H. Holland.
\newblock Genetic algorithms.
\newblock {\em Scientific American}, 267(1):66--73, 1992.

\bibitem{end_1}
J.~Andreoni.
\newblock Privately provided public goods in a large economy: The limits of
  altruism.
\newblock {\em Journal of Public Economics}, 35(1):57--73, 1988.

\bibitem{end_2}
Y.~Shi.
\newblock Economic description of tolerance in a society with asymmetric social
  cost functions.
\newblock {\em Economic research - Ekonomska istraživanja}, 31(1):2548--2593,
  2019.

\bibitem{end_3}
L.~Tesfatsion.
\newblock Agent-based computational economics: Growing economies from the
  bottom up.
\newblock {\em Artificial Life}, 8(1):55--82, 2002.

\bibitem{end_4}
W.~H. Marx, N.~L. DeMaintenon, K.~F. Mooney, M.~L. Mascia, J.~Medicis, P.~D.
  Franklin, E.~Sivak, and L.~Rotello.
\newblock Cost reduction and outcome improvement in the intensive care unit.
\newblock {\em The Journal of Trauma: Injury, Infection, and Critical Care},
  46(4):625--630, 1999.

\bibitem{end_5}
A.~H. Mutnick, K.~J. Sterba, J.~A. Peroutka, N.~E. Sloan, E.~A. Beltz, and
  M.~K. Sorenson.
\newblock {Cost savings and avoidance from clinical interventions}.
\newblock {\em American Journal of Health-System Pharmacy}, 54(4):392--396,
  1997.

\bibitem{end_6}
M.~Malach and W.J. Baumol.
\newblock {Further Opportunities for Cost Reduction of Medical Care}.
\newblock {\em Journal of Community Health}, 25:561--571, 2010.

\bibitem{intro_10}
M.~Ordu, E.~Demir, C.~Tofallis, and M.~M. Gunal.
\newblock A novel healthcare resource allocation decision support tool: A
  forecasting-simulation-optimization approach.
\newblock {\em Journal of the Operational Research Society}, 72(3):485--500,
  2021.

\bibitem{intro_9}
A.~Athanassopoulos and C.~Gounaris.
\newblock Assessing the technical and allocative efficiency of hospital
  operations in greece and its resource allocation implications.
\newblock {\em European Journal of Operational Research}, 133(2):416--431,
  2001.

\bibitem{intro_11}
J.~C. Lowery.
\newblock Simulations of a hospital's surgical suite and critical care area.
\newblock {\em Journal of the Operational Research Society}, 72(3):485--500,
  2021.

\end{thebibliography}
\bibliographystyle{unsrt}

\end{document}